# Temperature-dependent optical properties of plasmonic titanium nitride thin films

Harsha Reddy[1, 2], Urcan Guler[1, 2], Zhaxylyk Kudyshev[1], Alexander V. Kildishev[1], Vladimir M. Shalaev[1] and Alexandra Boltasseva[1,*]

[1]*School of Electrical & Computer Engineering and Birck Nanotechnology Center, Purdue University, West Lafayette, In 47907, USA*

[2]*These authors contributed equally to the work*

**ABSTRACT:** *Due to their exceptional plasmonic properties, noble metals such as gold and silver have been the materials of choice for the demonstration of various plasmonic and nanophotonic phenomena. However, noble metals' softness, lack of tailorability and low melting point along with challenges in thin film fabrication and device integration have prevented the realization of real-life plasmonic devices. In the recent years, titanium nitride (TiN) has emerged as a promising plasmonic material with good metallic and refractory (high temperature stable) properties. The refractory nature of TiN could enable practical plasmonic devices operating at elevated temperatures for energy conversion and harsh-environment industries such as gas and oil. Here we report on the temperature dependent dielectric functions of TiN thin films of varying thicknesses in the technologically relevant visible and near-infrared wavelength range from 330 nm to 2000 nm for temperatures up to 900 $^0$C using in-situ high temperature ellipsometry. Our findings show that the complex dielectric function of TiN at elevated temperatures deviates from the optical parameters at room temperature, indicating degradation in plasmonic properties both in the real and imaginary parts of the dielectric constant. However, quite strikingly, the relative changes of the optical properties of TiN are significantly smaller compared to its noble metal counterparts. In fact, at temperatures over 400 $^0$C the quality factors of localized surface plasmon resonances and propagating surface plasmons in thin TiN films become nearly the same as those in polycrystalline noble metals. Furthermore, no structural degradation was observed in any of TiN films upon heat treatment.*

*Using simulations, we demonstrate that incorporating the temperature-induced deviations into the numerical models leads to significant differences in the optical responses of high temperature nanophotonic systems. These studies hold the key for accurate modeling of high temperature TiN based optical elements and nanophotonic systems for energy conversion, harsh-environment sensors and heat-assisted applications.*

**INTRODUCTION:**

The last couple of decades have witnessed the emergence of plasmonics[1-3], optical metamaterials[4,5], and optical metasurfaces[6,7], which enable control over light at both the micro- and nanometer scale, beyond the limits of conventional optics. The ability to control the flow of light at the nanoscale has led to the demonstration of several intriguing phenomena ranging from subdiffraction imaging, cloaking, and negative index materials[8-10] to subwavelength waveguiding and light trapping for enhanced photovoltaic efficiencies[11,12]. Despite the significant progress, practical plasmonic applications have been limited. The major hurdle to the realization of practical plasmonic and metal based nanophotonic applications has been related to the properties of the constituent materials. The remarkable metallic properties of conventional noble metals, Au and Ag[13], have made them the natural choice of materials for investigating the plasmonic phenomena in the UV and visible parts of the spectrum. However, Au and Ag have relatively low bulk melting points around 1000 $^0$C[14,15], which further reduces when nanostructured due to melting point depression[16]. This precludes the usage of Au and Ag in plasmonic and nanophotonic applications imposing either high temperatures such as, heat assisted magnetic recording (HAMR)[17,18] and thermophotovoltaics, or high input intensities[15]. Noble metals are also incompatible with existing semiconductor fabrication process flow, preventing the practical realization of on-chip, CMOS-compatible applications[19]. These limitations have driven researchers to explore new materials for practical plasmonic applications[20,21].

One plasmonic material which has attracted a great deal of research interest in recent years is titanium nitride (TiN), a refractory transition metal nitride with a bulk melting point of 2930 $^0$C. It exhibits good metallic properties in the visible and near-IR parts of the spectrum, albeit inferior to Au and Ag at ambient temperature[20]. In addition, it is compatible with semiconductor process flow[14,20], unlike noble metals. It has been recently shown that the refractory nature of TiN helps in preserving the room temperature properties of a broadband metamaterial absorber upon heating to 800 $^0$C, where a similar design based on Au fails[15], revealing the advantages of TiN in high temperature nanophotonic applications. However, previous studies were limited to the room temperature optical response, while in-situ studies on the temperature evolution of TiN optical response have been largely missing. The temperature dependencies in 200 nm thick TiN films were considered previously. However, these studies were limited to a maximum temperature of 377 $^0$C and only 200 nm thick films were investigated[22]. Besides, recent studies on noble metals[23,24] have concluded that the temperature evolution of the optical properties is strongly dependent on the thickness of the film. Such a comprehensive study on the temperature dependent optical properties of TiN at higher temperatures, and varying film thicknesses has so far not been conducted.

Here, we report on the temperature induced deviations to the dielectric function of TiN thin films of varying thickness up to 900 $^0$C in the wavelength range from 330 nm to 2000 nm using a custom-built in-situ high temperature ellipsometry setup. We have investigated the temperature evolution of 30, 50 and 200 nm thick films deposited on sapphire substrates. We chose these thicknesses to make a comparison with the reports on Au and Ag, which considered films of similar thicknesses. Significant deviations in the dielectric function (both in the real and imaginary parts) were observed with increasing temperature. The real part decreases in magnitude, reducing the metallicity, while the imaginary part increases, making the films more lossy. At longer wavelengths ($\lambda > 1200 nm$), the real part at 900 $^0$C reduces by 40-50%, while the imaginary part increases by 30-45% compared to the room temperature values. Unlike their noble metal counterparts[23,24], no structural degradation was observed, while the surface morphology changes upon heat treatment.

We evaluated the plasmonic performance of TiN at high temperatures through numerical simulations and compared with that of Au and Ag. We considered a diabolo nanoantenna and investigated the influence of temperature dependent dielectric function on its opto-thermal response. We also provided the computed quality factors, performance metrics for plasmonics, using the extracted temperature dependent dielectric function. Quite remarkably, at temperatures over 400 °C the field enhancement and surface plasmon polariton quality factors in TiN become nearly the same as in Au and Ag polycrystalline films. We also provided causal analytical Drude-Lorentz oscillator models for the temperature dependent dielectric function.

**SAMPLE GROWTH AND EXPERIMENTAL APPROACH:**

The TiN samples were DC magnetron sputtered from a Ti target (Plasmaterials, 99.995%) onto (0001) sapphire substrate (GT Advanced Technologies, 10 mm x 10 mm x 0.5 mm) at 850 °C with a base pressure of $10^{-8}$ torr (200 W; 8 sccm N and 2 sccm Ar flow). Prior to the film growth the substrates were sonicated in acetone, isopropanol and methanol for 5 minutes each.

The temperature dependent studies were carried out in a custom built experimental platform consisting of a heating stage (Linkam TS1500) mounted onto a variable angle spectroscopic ellipsometer (VASE). Subsequently, dielectric functions at elevated temperatures were retrieved by fitting the VASE data with the Drude & 2 Lorentz oscillator models of the following form:

$$\hat{\varepsilon}(\omega) = \varepsilon_1 + i\varepsilon_2 = \varepsilon_\infty - \frac{\omega_p^2}{\omega^2 + i\Gamma_D \omega} + \sum_{j=1}^{2} \frac{\omega_{L,j}^2}{\omega_{0,j}^2 - \omega^2 - i\gamma_j \omega} \qquad (1)$$

where $\varepsilon_\infty$ is the background dielectric constant accounting for higher energy interband transitions outside the probed energy spectrum, and $\omega_p$ and $\Gamma_D$ are the plasma frequency and Drude damping, respectively. In addition, $\omega_{L,j}^2$, $\omega_{0,j}$, and $\gamma_j$ describe the Lorentz oscillator strength, energy, and damping, respectively. We use a two layer model consisting of sapphire/TiN layers while retrieving the temperature dependent dielectric function of TiN. We first collected room temperature ellipsometer data on sapphire substrate and extracted its optical properties by fitting with a Cauchy model. These properties were used for the sapphire layer while retrieving the temperature dependent dielectric function of

TiN. The Drude-Lorentz models of the temperature dependent dielectric function of TiN are available in the supporting information (Tables S1 - S3). More details about the experimental setup and modeling approach are available elsewhere[23,24].

**RESULTS:**

In Figure 1 we plot the experimentally extracted real ($\varepsilon_1$) and imaginary parts ($\varepsilon_2$) of the complex dielectric function of 200 nm ((a) & (b)), 50 nm ((c) & (d)) and 30 nm ((e) & (f)) TiN films at different temperatures, up to 900 $^0$C. The maximum temperature was limited to 900 $^0$C because of the experimental constraints but not due to the samples thermal stability. As the temperature is raised, both the real and imaginary parts show significant deviations from ambient temperature measurements. Particularly, the real part of all the films showed a monotonic decrease in their magnitude (Fig 1(a),(c),(e)), reducing the metallicity. Contrary to the conventional understanding, the observed decrease in the magnitude of the real part is primarily because of the increase in the Drude broadening $\Gamma_D$, but not due to the decrease in the plasma frequency $\omega_p$ [24]. In fact, $\omega_p$ marginally increases with temperature up to 800 $^0$C (maximum deviation < 6%, see Drude-Lorentz oscillator models in Tables S1-S3). The observed temperature dependence in $\varepsilon_1$ can be understood from the drude ter in Eq (1), which captures the free electron response:

$$\hat{\varepsilon} = \varepsilon_1 + i\varepsilon_2 = \frac{-\omega_p^2}{\omega^2 + i\Gamma_D \omega} \qquad (2)$$

$$\text{where } \varepsilon_1 = \frac{-\omega_p^2}{\omega^2 + \Gamma_D^2} \text{ and } \varepsilon_2 = \frac{\omega_p^2 \Gamma_D}{\omega^3 + \Gamma_D^2 \omega}. \qquad (3)$$

when $\Gamma_D \ll \omega$, which is the case with noble metals close to ambient temperatures, the real part can be approximated as $\varepsilon_1 \approx \frac{-\omega_p^2}{\omega^2}$, independent of $\Gamma_D$. This approximation is no longer valid when $\Gamma_D$ is comparable to $\omega$. For TiN films $\Gamma_D$ varies from 0.2 - 0.75 eV depending on the temperature and film thickness (Tables S1-S3 in supporting information). For the photon energy ranges considered in the current study, particularly at near-IR wavelengths, the above approximation is no longer valid. From the exact expression of $\varepsilon_1$ in Eq (3) it is clear that an increase in $\Gamma_D$ leads to a decrease in the magnitude of $\varepsilon_1$, which is what we observed in all TiN films. A similar evolution of $\varepsilon_1$ with temperature was reported in a recent study on Ag[24].

The imaginary part at longer wavelengths on the other hand, increases with temperature up to 800 $^0$C (Fig 1(b),(d),(f)). Increasing the temperature even further leads to a decrease in $\varepsilon_2$ (see Fig 2b). The observed deviations in the imaginary part are due to the increase in $\Gamma_D$ and the non-monotonic behavior of $\omega_p$. As pointed out earlier, $\Gamma_D$ increases monotonically with temperature. The plasma frequency on the other hand deviates marginally with temperature, increasing until 800 $^0$C and decreasing when the temperature is increased to 900 $^0$C (see Tables S1-S3 in supporting information). As a result, the numerator in the Drude expression of $\varepsilon_2$ in Eq (3) $\omega_p^2 \Gamma_D$, increases monotonically, increasing the imaginary part. At the same time, the denominator $\omega^3 + \Gamma_D^2 \omega$, increases quadratically with $\Gamma_D$, counteracting the increase in the numerator. Based on the experimental evidence, we conclude that the increase in $\omega_p^2 \Gamma_D$ dominates until 800 $^0$C. However, at 900 $^0$C the quadratic dependence of the denominator on $\Gamma_D$, along with the marginal decrease in $\omega_p$, lead to a decrease in the imaginary part.

The observed temperature dependencies in $\Gamma_D$ and $\omega_p$ are predominantly due to three physical mechanisms: 1) increasing electron-phonon scattering, 2) changing electron effective mass, and 3) reducing carrier densities. Qualitatively, the increase in electron-phonon interaction can be understood by noting that the phonon number increases with temperature, thereby increasing the scattering rates and $\Gamma_D$. The plasma frequency on the other hand depends on the free electron density $N$ and the effective mass $m^*$ as:

$$\omega_p^2 = \frac{Ne^2}{m^* \varepsilon_0} \quad (4)$$

As the temperature is raised, $N$ reduces due to volume thermal expansion, decreasing the plasma frequency. The effective mass on the other hand, is known to reduce with temperature[32]. The interplay between these two counteracting mechanisms leads to the observed temperature dependencies in $\omega_p$. From the experimentally extracted plasma frequency (Fig S4, Tables S1-S3) we conclude that the decrease in $m^*$ is the dominant mechanism until 800 ⁰C, while decrease in $N$ dominates at 900 ⁰C. Similar increasing and decreasing trends in $\omega_p$ were observed in Au[23] and Ag[24]. We would like to point out that the relative deviations in the plasma frequency are only marginal and don't have a significant bearing on the evolution of the optical properties. Figure 2 shows the deviations in the real and imaginary parts as a function of temperature at 1550 nm wavelength. More details on the temperature dependencies are available in earlier works[23,24].

We would like to point out a couple of key differences between the evolution of optical properties in TiN compared to those in noble metals, Au and Ag, reported in earlier studies[23,24]. In thin Au and Ag films, both $\varepsilon_1$ and $\varepsilon_2$ showed significantly larger deviations with temperature compared to their thicker counterparts (nearly fourfold increase in $\varepsilon_2$ and nearly 30 % decrease the magnitude of $\varepsilon_1$ in thin films compared to twofold increase in $\varepsilon_2$ and 10 - 15% decrease in the magnitude of $\varepsilon_1$ in thick films). In contrast to the noble metals, the temperature evolution of the dielectric function in TiN is nearly independent of the thickness. Furthermore, thin Au and Ag polycrystalline (single crystalline) films showed severe structural and surface morphological degradation upon heating to temperatures close to 400 ⁰C (600 ⁰C)[23,24]. However, no such structural or morphological degradation was observed in TiN films, even though the samples were heated to higher temperatures (AFM topographs on 50 nm TiN are shown in Fig S5).

We estimated the plasmonic performance of TiN at high temperatures using quality factors, plasmonic performance metrics, for both localized and propagating plasmons[25,26]. For localized surface plasmons, the strength of the local electric field enhancement in a spherical nanoparticle, under quasi-static assumption, is given by $Q_{LSPR} = \frac{-\varepsilon_1}{\varepsilon_2}$ [25]. Whereas for propagating plasmons, the decay length of surface plasmon polaritons ($L_{SPP}$), serves as a figure of merit for waveguiding applications[26]. In Figure 3 we plot the quality factors of the localized surface plasmon resonances and the propagation lengths of surface plasmon resonances, computed using the temperature dependent dielectric functions from 30 nm TiN. The quality factors of 50 nm and 200 nm TiN films are given in supporting information Fig S1 & S2. Figure 4 shows a comparison of the quality factors of 30 nm Au and Ag (taken from Ref 22 and 23), along with 30 nm TiN, at various temperatures. From Figure 4 it is clear that although TiN has field enhancement and surface plasmon polaritons (SPP) propagation quality factors inferior to noble metals at room temperature, the difference between the performance metrics reduces substantially and the quality factors in TiN approach those in Au and Ag at temperatures over 400 ⁰C. Table 1 (2) below shows a comparison of $Q_{LSPR}$ (SPP propagation lengths) in TiN, Au and Ag at different temperatures and a wavelength of 1550 nm.

We also considered the role of temperature induced deviations in generating nanoscale heat sources. Absorption and losses in metals have long been considered to be detrimental to plasmonic performance. However, in the emerging field of thermoplasmonics, the goal is to take advantage of the strong optical absorption in plasmonic nanostructures and generate nanoscale thermal sources[27-30]. These nanoscale thermal sources are of great interest in a wide variety of applications such as cancer therapy, photo-thermal imaging, nanochemistry and opto/plasmo fluidics[27]. By engineering the geometries of nanoantennas, temperatures as high as 800 ⁰C can be generated locally in plasmonic components (with continuous wave excitation), while keeping the background/substrate close to room temperature. These intense nanoscale heat sources can be of great use in chemical reactions requiring high temperatures, such as ammonia synthesis, methane oxidation and several other heterogeneous chemical reactions, alleviating the need for bulky experimental setups operating at very high temperatures[31].

Here we show that incorporating the temperature induced deviations would lead to significantly different opto-thermal responses in plasmonic heat sources. We performed a coupled opto-thermal finite-element method (FEM) based multiphysics simulation of heating in diabolo antenna (DA) arrays, induced by external optical excitation. We numerically considered three different cases of DA arrays on 0.5 mm thick sapphire substrate: made of gold, silver and TiN, with optical resonances at 1.14 µm, 1.16 µm and 1.3 µm, respectively. Dimensions of the unit cell are shown on Fig. 5a. We assume that the diabolo antennas are illuminated with a Gaussian beam profile with power $P_{in}$, and a full width at half maximum of 19.5 µm at resonance wavelengths exciting 936 antennas.

Heat generation in plasmonic structures is governed by Joule heating, with the strength of the heat source determined by dissipated electromagnetic power density inside the particle. The spatial heat source density is determined by the distribution of the local electromagnetic inside the plasmonic DA, as well as by the imaginary part of dielectric function as:

$$q(r,T) = \varepsilon_0 \omega \operatorname{Im}(\varepsilon(T))\left|\vec{E}(r)\right|^2$$

where $\varepsilon_0$, $\omega$ and $\vec{E}(r)$ are the permittivity of free space, frequency of the incident photons and local electric field, respectively, while $\varepsilon(T)$ is the temperature dependent dielectric function of the metal. The dissipated energy is determined by both the distribution of the local electric field inside the plasmonic DA, as well as the imaginary part of the dielectric function. However, since the imaginary part has a strong dependence on temperature, estimating the temperature rise in the DA requires a recursive calculation that takes into account the temperature dependence of $\varepsilon$. We performed these calculations using the experiment fitted temperature dependent dielectric functions of 30 nm Au, 40 nm Ag (taken from Ref 21 & 22) and 30 nm TiN films. The coupled opto-thermal simulations were performed using a commercial finite element method based solver (COMSOL Multiphysics). A typical spatial distribution of the dissipated power density in a diab0lo antenna at resonant condition is shown in Fig. 5b. In

Figure 5c we show the dependence of the maximum temperature on the surface of diabolo nanoantenna as a function of input power $P_{in}$, computed using the temperature induced deviations. For a comparison, the expected temperature rise using the room temperature imaginary part is also plotted. These results clearly show that incorporating the temperature dependencies lead to significantly higher temperatures in the plasmonic thermal sources, particularly for high input powers. For instance, at the maximum incident powers considered here, nearly 41%, 145% and 27% increase in the maximum temperature is expected in Au, Ag and TiN, respectively. We would like to point out that despite pumping with higher powers the TiN based DA has a more stable response due to its refractory nature, exhibiting smaller relative deviations in the maximum temperature, compared to their noble metal counterparts. Fig 5d shows the computed temperature field in TiN DA when excited with 61 mW input power. The temperature field is nearly uniform because of the high thermal conductivity of metals, unlike the electric field which exhibits strong spatial localization.

**CONCLUSION:**

We have experimentally extracted the temperature dependent optical properties of plasmonic TiN thin films of 200, 50 and 30 nm TiN films on sapphire in the wavelength range 330- 2000 nm up to 900 $^0$C. Our findings show that both the real and imaginary parts of the complex dielectric function change considerably with the temperature. The remarkable changes in the optical properties lead to deviations in the quality factors for localized surface plasmons as well as the propagation lengths of SPPs. However, the relative decay in the TiN quality factors are significantly smaller compared to Au and Ag (60-65% decrease in the TiN quality factors compared with 80% decrease in Au and Ag, at the largest temperatures). Although the noble metals exhibit superior plasmonic performance compared to TiN at room temperature, at temperatures over 400 $^0$C the performance of TiN approaches that in Au and Ag, with TiN quality factors becoming nearly the same as in Au and Ag. In addition, no structural degradation was observed in TiN films. These findings demonstrate the potential of TiN for high temperature nano-photonic applications. We demonstrate that including the temperature dependencies into numerical models lead to significantly different optical responses. We therefore provide causal analytical models that describe the temperature dependent dielectric functions. These analytical models will be of critical importance for the design and optimization of TiN based nanophotonic and plasmonic components for practical high temperature applications for energy conversion, harsh-environment sensors and heat-assisted concepts in data recording, imaging and photothermal therapy.

**AUTHOR INFORMATION**

**Corresponding Author**

*E-mail: aeb@purdue.edu

**Notes**

The authors declare no competing financial interest.

**ACKNOWLEDGEMENTS**

This work was supported in part from NSF OP (DMR-1506775), and NSF MRSEC (DMR-1120923). The authors would like to acknowledge help from Prof. Ali Shakouri, and Yee Rui Koh with the experimental setup.

*Solids,* **1969**, 30, 2765–2769.

**Figures:**

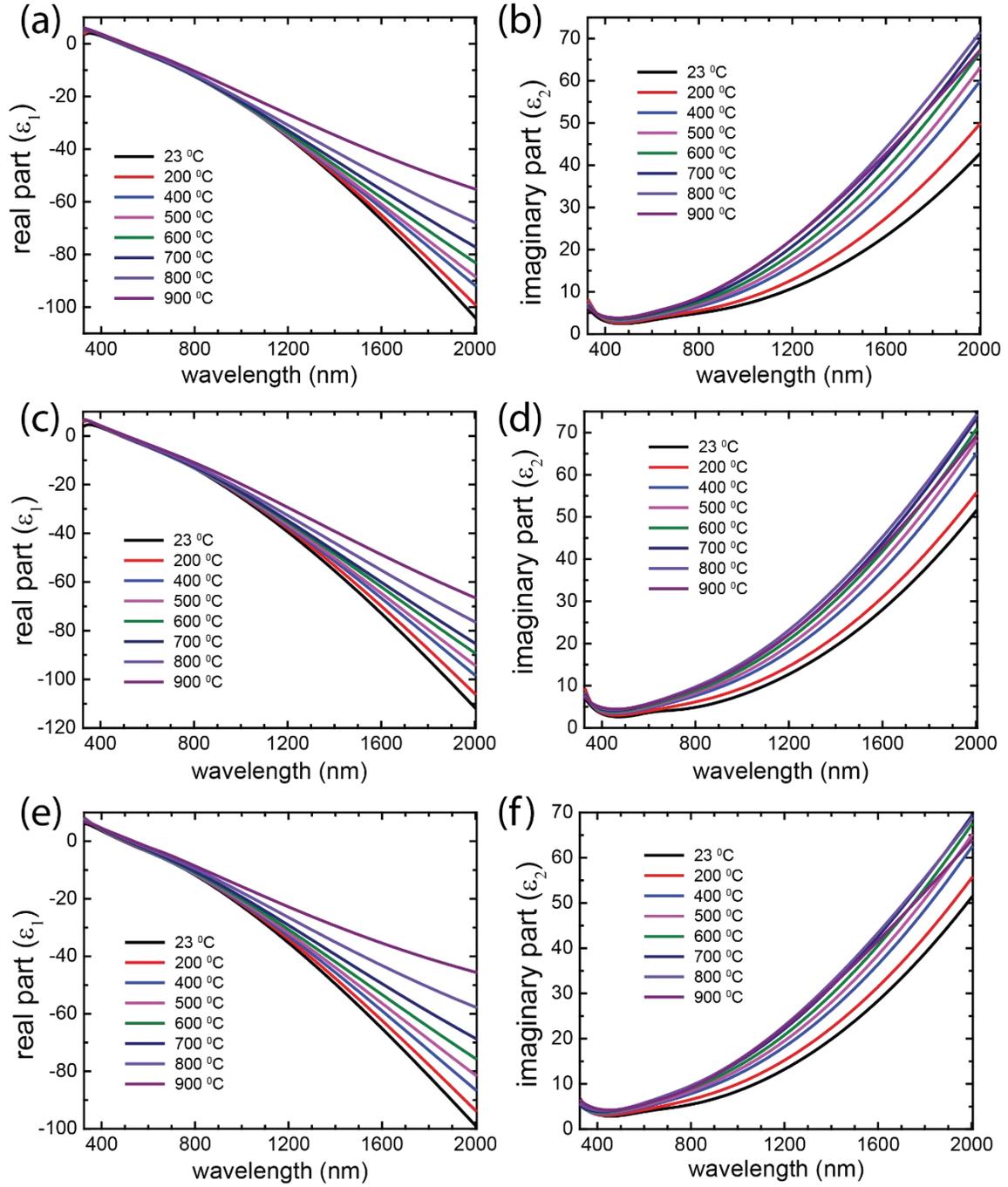

**Fig 1: Temperature dependent dielectric functions of 200 nm (a, b), 50 nm (c, d) and 30 nm (e, f) TiN films.** With increasing temperature the real part decreases in magnitude, while the imaginary part increases up to 800 °C. Increasing the temperature even further reduces the imaginary part. The color coding is shown in the legend.

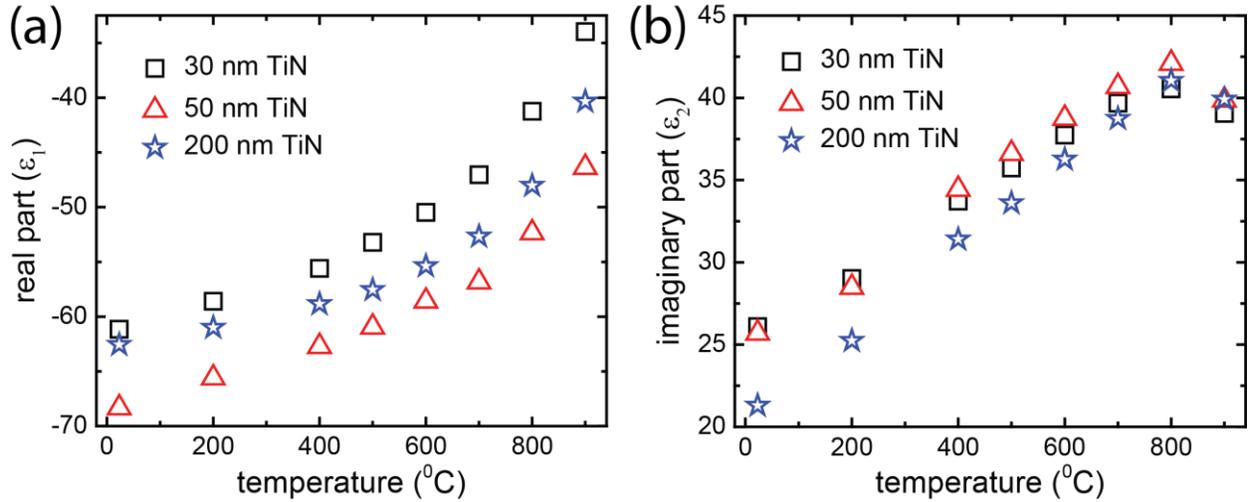

**Fig 2: Real (a) and imaginary parts (b) of the complex dielectric function at a wavelength of 1550 nm as a function of temperature.** The real part shows a monotonic increase with temperature. The imaginary part increases till 800 °C and raising the temperature even further reduces it.

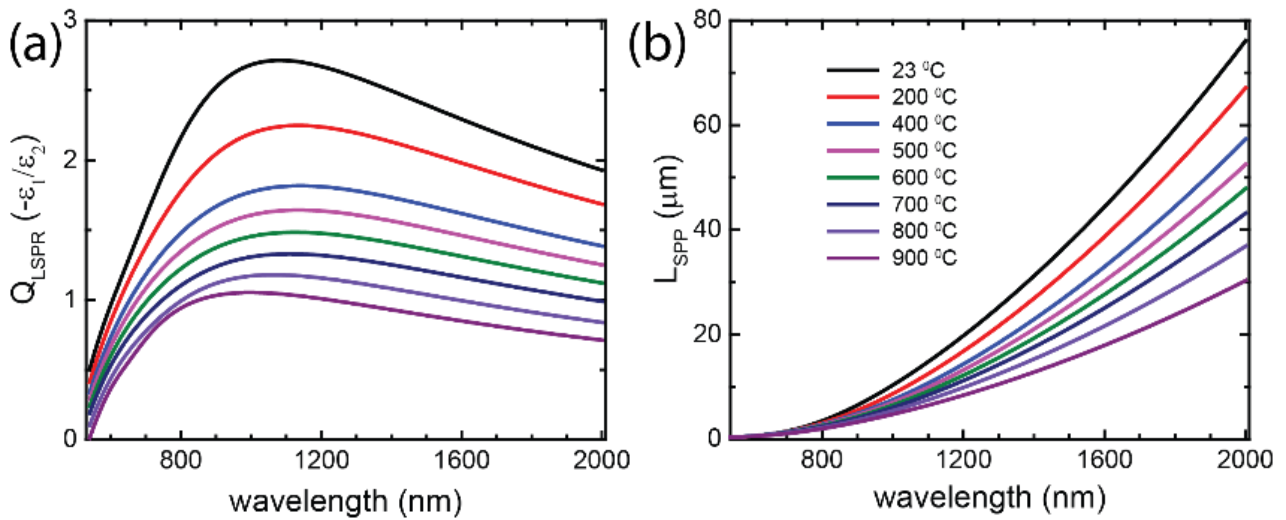

**Fig 3: Temperature dependent quality factors of localized surface plasmon resonances (a) and propagation lengths of SPPs at the metal-air interface (b) computed with optical constants from 30 nm TiN film.** Both $Q_{LSPR}$ and $L_{SPP}$ decrease with increasing temperature. Legend in (b) shows the color coding for both the plots.

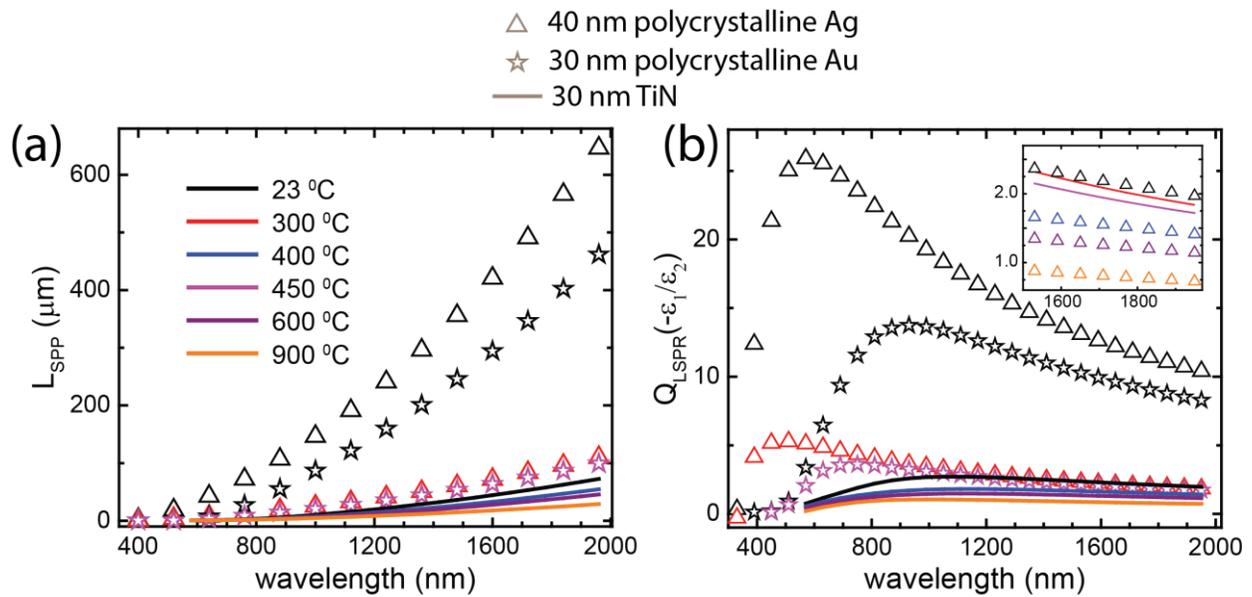

**Fig 4: Comparison of Au, Ag and TiN plasmonic performance at high temperatures.** SPP propagation lengths (a) and $Q_{LSPR}$ (b) of 40 nm Ag (open triangles), 30 nm Au (open stars) and 30 nm TiN (solid lines) at different temperatures. Legend in (a) shows the temperature color coding for both the plots. The $Q_{LSPR}$ at longer wavelengths are shown in the inset of (b).

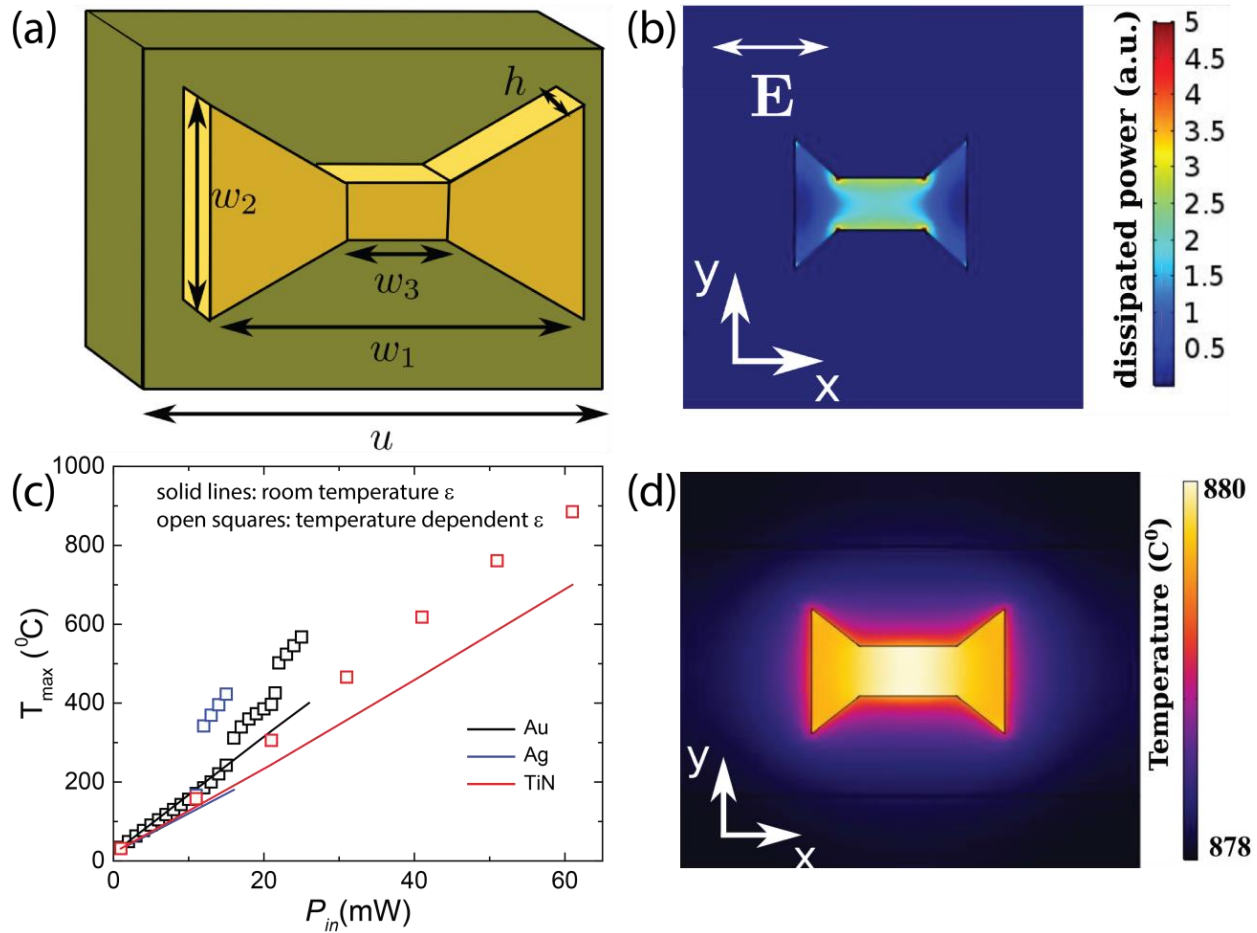

**Fig 5: Opto-thermal response of a diabolo antenna.** (a) Diabolo unit cell u = 500 nm, $w_1$ = 235 nm, $w_2$ = 150 nm, $w_3$ = 100 nm, h= 50 nm. (b) Typical spatial distribution of dissipated power in a diabolo antenna. (c) Maximum temperature on the diabolo surface as a function input power in Au, Ag and TiN. Solid lines correspond to the temperature rise with room temperature dielectric function, while open squares correspond to the temperature rise when the temperature dependencies are included into FEM models. (d) Spatial profile of temperature in a TiN diabolo antenna, computed using the temperature deviations with 61 mW input power. The temperature profile is nearly uniform across the nanoantenna, unlike the strongly localized electric field.

| Sample type | $Q_{LSPR}$ at 23 °C | $Q_{LSPR}$ at 300 °C | $Q_{LSPR}$ at 400 °C | $Q_{LSPR}$ at 450 °C | $Q_{LSPR}$ at 600 °C | $Q_{LSPR}$ at 900 °C | % change at the largest temperature |
|---|---|---|---|---|---|---|---|
| 30 nm PC Au | 10.2 | 10.2 | 4.9 | 2.1 | - | - | 79.4 |
| 40 nm PC Ag | 12.9 | 2.3 | - | - | - | - | 82.2 |

| | | | | | | |
|---|---|---|---|---|---|---|
| 30 nm TiN | 2.3 | - | 1.6 | - | 1.3 | 0.87 | 62.2 |
| 50 nm TiN | 2.7 | - | 1.8 | - | 1.5 | 1.2 | 55.5 |
| 200 nm TiN | 2.9 | - | 1.9 | - | 1.5 | 1.01 | 65.2 |

**Table 1: Computed quality factors of localized surface plasmon resonances in Au, Ag and TiN at various temperatures and a wavelength of 1550 nm.**

| Sample type | $L_{SPP}$ in $\mu m$ at 23 °C | $L_{SPP}$ in $\mu m$ at 300 °C | $L_{SPP}$ in $\mu m$ at 400 °C | $L_{SPP}$ in $\mu m$ at 450 °C | $L_{SPP}$ in $\mu m$ at 600 °C | $L_{SPP}$ in $\mu m$ at 900 °C | % change at the largest temperature |
|---|---|---|---|---|---|---|---|
| 30 nm PC Au | 273 | 265 | 136 | 59 | - | - | 78.4 |
| 40 nm PC Ag | 393 | 67 | - | - | - | - | 82.9 |
| 30 nm TiN | 41 | - | 30 | - | 25 | 17 | 58.5 |
| 50 nm TiN | 55 | - | 36 | - | 31 | 23 | 58.2 |
| 200 nm TiN | 50 | - | 34 | - | 29 | 20 | 60 |

**Table 2: Computed propagation lengths of SPPs in Au, Ag and TiN at various temperatures and a wavelength of 1550 nm.**

## Supporting Information

We use the following form of Drude-Lorentz model to describe the dielectric function:

$$\hat{\varepsilon}(\omega) = \varepsilon_1 + i\varepsilon_2 = \varepsilon_\infty - \frac{\omega_p^2}{\omega^2 + i\Gamma_D \omega} + \sum_{j=1}^{2} \frac{\omega_{L,j}^2}{\omega_{0,j}^2 - \omega^2 - i\gamma_j \omega}$$

The temperature dependencies of the plasma frequency $\omega_p$ and Drude broadening $\Gamma_D$ were discussed in previous reports (Ref 23 in the manuscript). Briefly, the observed temperature deviations are due to the changing effective mass, reducing carrier densities and increasing electron phonon interactions. With increasing temperature, $\Gamma_D$ increases monotonically. Intuitively, this monotonic increase can be understood by noting that the phonon number, which follows Bose-Einstein statistics, increases with temperature. The larger phonon number leads to larger electron-phonon interactions, decreasing the scattering times and thereby increasing the Drude broadening. As discussed in the earlier works on noble metals, $\Gamma_D$ varies linearly with temperature, as long as the microstructure of the films remains intact. A similar temperature dependence was observed in TiN films. The plasma frequency on the other hand increases until 800 $^0$C. Raising the temperature further leads to a decrease in the plasma frequency. We attribute this increasing and decreasing trends to the interplay between the reducing electron effective mass and carrier density (see ref 23 in the manuscript for additional details). In Fig S3 we plot the plasma frequency and Drude broadening of 200 nm TiN as a function of temperature. The Drude broadening was fitted with a linear temperature dependence.

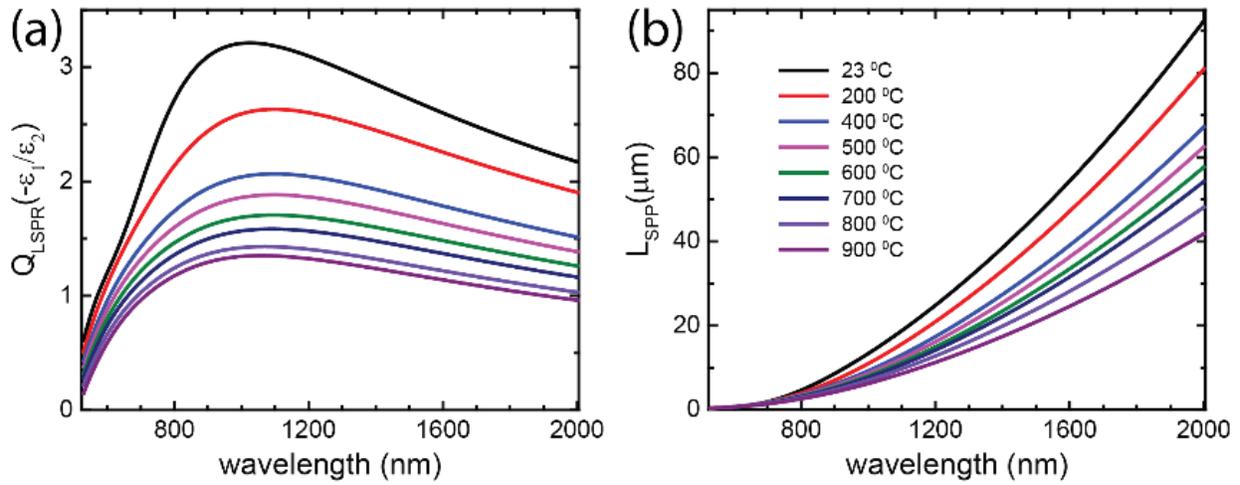

**Fig S1: Temperature dependent quality factors of localized surface plasmon resonances (a) and propagation lengths of SPPs at the metal-air interface (b) computed with optical constants from 50 nm TiN film.** Both $Q_{LSPR}$ and $L_{SPP}$ decrease with increasing temperature. Legend in (b) shows the color coding for both plots.

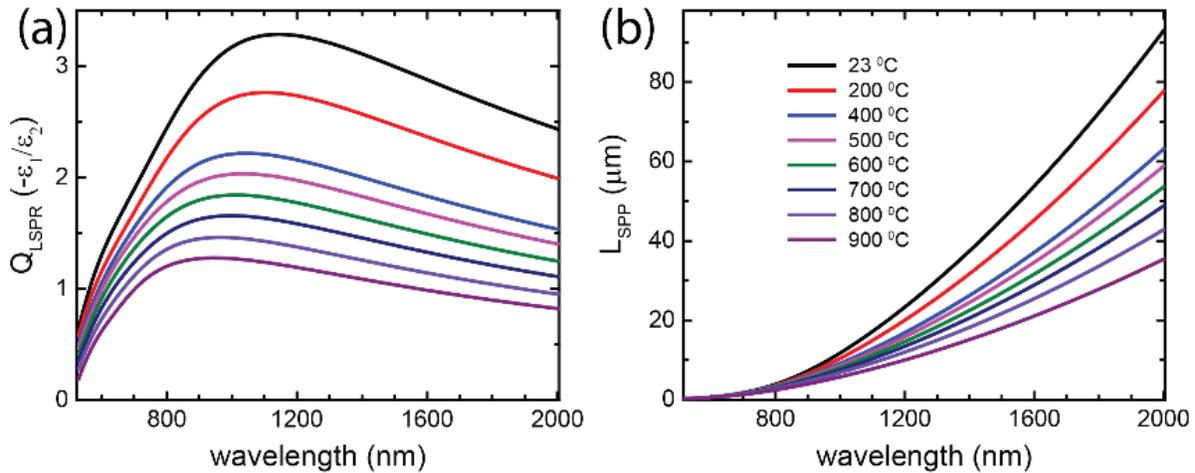

**Fig S3: Temperature dependent quality factors of localized surface plasmon resonances (a) and propagation lengths of SPPs at the metal-air interface (b) computed with optical constants from 200 nm TiN film.** Both $Q_{LSPR}$ and $L_{SPP}$ decrease with increasing temperature. Legend in (b) shows the color coding for both plots.

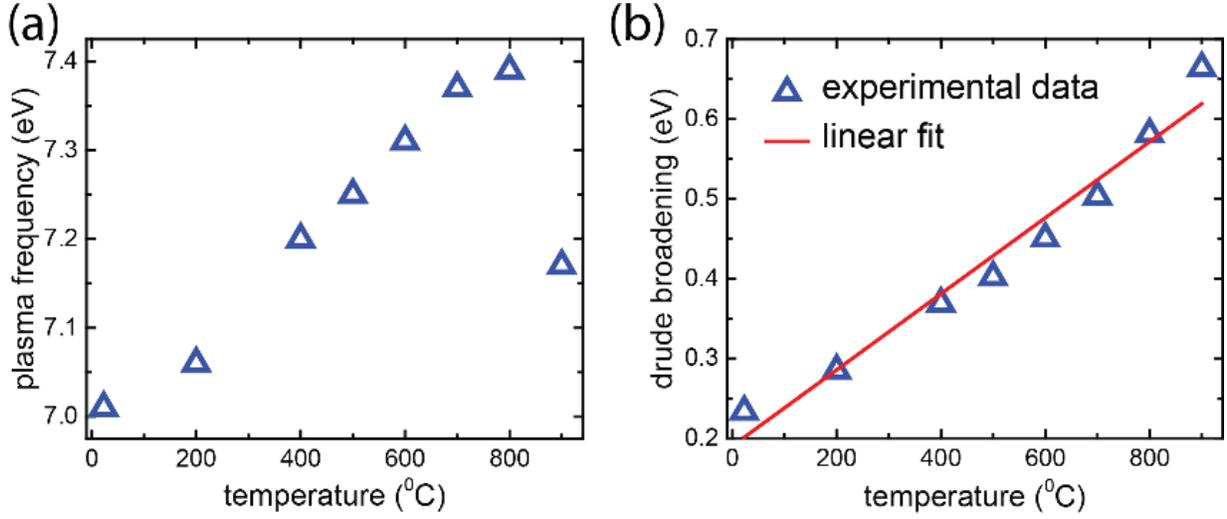

**Fig S4: Plasma frequency (a) and drude broadening (b) of 200 nm TiN film as a function of temperature.** The plasma frequency varies marginally with temperature, increasing till 800 $^0$C and then reduces upon increasing the temperature to 900 $^0$C. The drude broadening increases monotonically with temperature. A linear fit describes the experimental drude broadening well.

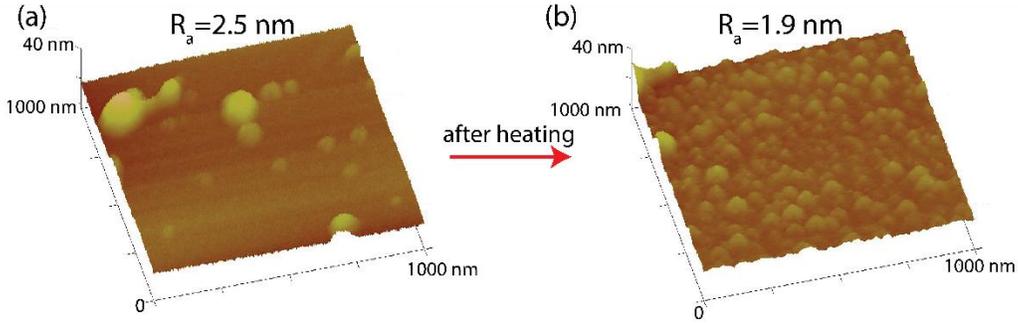

**Fig S5: AFM topographs from before (a) and after (b) heat treatment of 50 nm TiN.**

| $T(^0C)$ | $\varepsilon_\infty$ | $\omega_p(eV)$ | $\Gamma_D(eV)$ | $\omega_{L,1}^2(eV^2)$ | $\gamma_1(eV)$ | $\omega_{0,1}(eV)$ | $\omega_{L,2}^2(eV^2)$ | $\gamma_2(eV)$ | $\omega_{0,2}(eV)$ |
|---|---|---|---|---|---|---|---|---|---|
| 23  | 4.74 | 7.01 | 0.23 | 36.7  | 1.45 | 4.1  | 3.88 | 1.33 | 1.85 |
| 200 | 4.35 | 7.06 | 0.28 | 39    | 1.15 | 4.01 | 4.33 | 1.38 | 1.92 |
| 400 | 3.89 | 7.20 | 0.37 | 54.4  | 1.40 | 4.24 | 3.61 | 1.46 | 1.95 |
| 500 | 3.58 | 7.25 | 0.40 | 66.3  | 1.54 | 4.39 | 3.58 | 1.54 | 1.94 |
| 600 | 3.37 | 7.31 | 0.45 | 76.8  | 1.70 | 4.50 | 2.91 | 1.44 | 1.96 |
| 700 | 3.14 | 7.37 | 0.50 | 92.2  | 1.94 | 4.67 | 2.22 | 1.27 | 1.94 |
| 800 | 2.67 | 7.39 | 0.58 | 114.9 | 2.13 | 4.90 | 1.15 | 0.92 | 2.00 |
| 900 | 2.08 | 7.17 | 0.66 | 122.6 | 2.13 | 4.96 | 0.46 | 0.58 | 2.01 |

**Table S1: Temperature dependent Drude + Lorentz models of 200 nm TiN.**

| $T(^0C)$ | $\varepsilon_\infty$ | $\omega_p(eV)$ | $\Gamma_D(eV)$ | $\omega_{L,1}^2(eV^2)$ | $\gamma_1(eV)$ | $\omega_{0,1}(eV)$ | $\omega_{L,2}^2(eV^2)$ | $\gamma_2(eV)$ | $\omega_{0,2}(eV)$ |
|---|---|---|---|---|---|---|---|---|---|
| 23  | 5.18 | 7.38 | 0.26 | 42.2 | 1.42 | 4.07 | 2.26 | 0.87 | 2.02 |
| 200 | 4.43 | 7.37 | 0.30 | 47.9 | 1.04 | 4.09 | 5.76 | 1.75 | 1.99 |
| 400 | 4.28 | 7.49 | 0.37 | 60.0 | 1.19 | 4.26 | 7.05 | 2.03 | 1.93 |
| 500 | 4.28 | 7.52 | 0.40 | 66.0 | 1.30 | 4.33 | 7.29 | 2.06 | 1.90 |
| 600 | 4.30 | 7.55 | 0.43 | 70.3 | 1.40 | 4.37 | 7.68 | 2.14 | 1.86 |
| 700 | 4.06 | 7.59 | 0.47 | 81.7 | 1.54 | 4.47 | 7.11 | 2.00 | 1.86 |
| 800 | 3.21 | 7.55 | 0.53 | 110 | 1.83 | 4.72 | 5.38 | 1.82 | 1.84 |
| 900 | 1.00 | 7.27 | 0.57 | 187 | 2.30 | 5.23 | 3.01 | 1.49 | 1.78 |

**Table S3: Temperature dependent Drude + Lorentz models of 50 nm TiN.**

| $T(^0C)$ | $\varepsilon_\infty$ | $\omega_p(eV)$ | $\Gamma_D(eV)$ | $\omega_{L,1}^2(eV^2)$ | $\gamma_1(eV)$ | $\omega_{0,1}(eV)$ | $\omega_{L,2}^2(eV^2)$ | $\gamma_2(eV)$ | $\omega_{0,2}(eV)$ |
|---|---|---|---|---|---|---|---|---|---|
| 23  | 3.93 | 7.10 | 0.29 | 62.9 | 1.39 | 4.40 | 3.97 | 1.35 | 2.03 |
| 200 | 3.75 | 7.12 | 0.33 | 65.3 | 0.97 | 4.42 | 8.53 | 2.22 | 2.08 |
| 400 | 3.17 | 7.21 | 0.39 | 95.9 | 1.15 | 4.77 | 10.53 | 2.59 | 2.00 |
| 500 | 2.64 | 7.22 | 0.43 | 125.4 | 1.37 | 5.04 | 9.80 | 2.48 | 1.95 |
| 600 | 2.16 | 7.25 | 0.48 | 158.8 | 1.63 | 5.30 | 8.34 | 2.22 | 1.90 |
| 700 | 1.30 | 7.28 | 0.54 | 211 | 1.92 | 5.62 | 6.61 | 1.94 | 1.88 |
| 800 | 1.00 | 7.24 | 0.63 | 227.8 | 2.29 | 5.68 | 2.98 | 1.34 | 1.90 |
| 900 | 1.00 | 7.04 | 0.74 | 197.7 | 2.57 | 5.47 | 0.31 | 0.46 | 1.96 |

**Table S2: Temperature dependent Drude + Lorentz models of 30 nm TiN.**